\documentclass[prd,aps,a4,onecolumn,superscriptaddress,preprintnumbers,nofootinbib]{revtex4-1}

\usepackage{pslatex}
\usepackage[pdftex]{graphicx}
\usepackage{psfrag}
\usepackage{epsfig}
\usepackage{color}
\usepackage{cancel}
\usepackage{slashed}
\usepackage{amssymb}
\usepackage{amsmath}
\usepackage{hyperref}
\usepackage{enumerate}
\usepackage{multirow}
\usepackage{ulem}
\usepackage{float}
\usepackage{comment}
\usepackage{diagbox}
\usepackage{amsmath}
\usepackage{soul, xcolor}
\usepackage{makecell}

\usepackage{calligra}
\DeclareMathAlphabet{\mathcalligra}{T1}{calligra}{m}{n}
\DeclareFontShape{T1}{calligra}{m}{n}{<->s*[2.2]callig15}{}
\newcommand{\scriptr}{\mathcalligra{r}\,}

\def\bse{\begin{eqnarray*}}
	\def\ese{\end{eqnarray*}}
\def\pr{\hbox{pr}}

\def\D{\mathcal{D}}
\def\T{\mathcal{T}}
\allowdisplaybreaks

\begin{document}
\setstcolor{red}

\title{Detection of astrophysical neutrinos at prospective locations of dark matter detectors}

\author{Yi Zhuang}
\affiliation{Department of Physics and Astronomy, Mitchell Institute
  for Fundamental Physics and Astronomy, Texas A\&M University,
  College Station, Texas 77843, USA}
\author{Louis E. Strigari}%
\affiliation{Department of Physics and Astronomy, Mitchell Institute
  for Fundamental Physics and Astronomy, Texas A\&M University,
  College Station, Texas 77843, USA}

\author{Lei Jin}%
\affiliation{Department of Mathematics and Statistics, Texas A\&M University-Corpus Christi,  Corpus Christi,  TX 78412, USA}

\author{Samiran Sinha}%
\affiliation{Department of Statistics, Texas A\&M University,
  College Station, Texas 77843, USA}
  
\date{\today}
\begin{abstract}
We study the prospects for detection of solar, atmospheric neutrino and DSNB fluxes at future large-scale dark matter detectors through both electron and nuclear recoils. We specifically examine how the detection prospects change for several prospective detector locations (SURF, SNOlab, Gran Sasso, CJPL, and Kamioka), and improve upon the statistical methodologies used in previous studies. Due to its ability to measure lower neutrino energies than other locations, we find that the best prospects for the atmospheric neutrino flux are at the SURF location, while the prospects are weakest at CJPL because it is restricted to higher neutrino energies. On the contrary, the prospects for the diffuse supernova neutrino background (DSNB) are best at CJPL, due largely to the reduced atmospheric neutrino background at this location. Including full detector resolution and efficiency models, the CNO component of the solar flux is detectable via the electron recoil channel with exposures of $\sim 10^3$ ton-yr for all locations. These results highlight the benefits for employing two detector locations, one at high and one at low latitude. 
\end{abstract}


\maketitle

\section{Introduction \label{sec:introduction}}

\par Over the past several decades, direct dark matter detection experiments have made tremendous progress in constraining weak-scale particle dark matter~\cite{XENON:2018voc,LUX-ZEPLIN:2022qhg,2014PandaX,2009CDMS}. Future larger-scale detectors will be sensitive to not only particle dark matter, but also astrophysical neutrinos and various other rare-event phenomenology~\cite{Aalbers:2022dzr,2014Grothaus,Spergel1988,2012Billard}. The most prominent of the neutrino signals are from the Sun, the atmosphere, and the diffuse supernova neutrino background (DSNB)~\cite{1989Bahcall,Billard:2013qya,2010Beacom,1999Scholberg}. Understanding these signals has important implications for the future of particle dark matter searches, and also for understanding the nature of the sources and the properties of neutrinos~\cite{Dutta:2019oaj}. 

\par Various methods have been proposed to distinguish neutrinos and a possible dark matter signal. These include exploiting the energy distribution of nuclear recoils between neutrinos and dark matter~\cite{Dent:2016iht,Dent:2016wor}, the differences in arrival directions~\cite{OHare:2015utx}, and the differences in the periodicities of the signal~\cite{Davis:2014ama}. New physics in the neutrino sector may also change the nature of the predicted neutrino signal~\cite{Cerdeno:2016sfi,AristizabalSierra:2019ykk}, and provide a method to discriminate from dark matter. 

\par In this paper, we examine the prospects for detecting all of these neutrino flux components at large-scale, next generation detectors. We consider detection through both the electron recoil and the nuclear recoil channels. We present a principled statistical methodology for extracting all flux components, and compare it to to previous methods that attempted to extract some of the flux components that we consider~\cite{2019Jayden,2021Jayden}. We mainly focus on how the detection prospects for all flux components depend on detector location, considering 5 specific detectors locations: China Jinping Underground Laboratory (CJPL), Kamioka, Laboratori Nazionali del Gran Sasso (LNGS), SNOlab and the Sanford Underground Research Facility (SURF). The location of the detector is crucial because the atmospheric neutrino flux is from cosmic ray interacting with atmosphere, where primary cosmic ray is cut off by geomagnetic field which depends strongly on the detector's latitude and longitude~\cite{1983Cooke,Zhuang:2021rsg}. Including this effect is important not only for detecting the atmospheric neutrino flux itself but also for other subdominant components, such as the DSNB, for which the atmospheric neutrino component is a background. 

\par This paper is organized as follows. In Section~\ref{sec:SandB}, we describe the nature of the signals and backgrounds that we use in our analysis. In Section~\ref{sec:NEST}, we describe the simulations of detector properties to interpret the signals. In Section~\ref{sec:statistical}, we introduce the statistical methodologies used in our analysis, and compare to previous analysis methods. Then in Section~\ref{sec:results}, we present our resulting projections, and in Section~\ref{sec:conclusions}, the discussion and conclusions. 

\section{Signal and Backgrounds \label{sec:SandB}}
\par
\begin{figure*}[!htbp]
\includegraphics[width = 0.95\textwidth ]{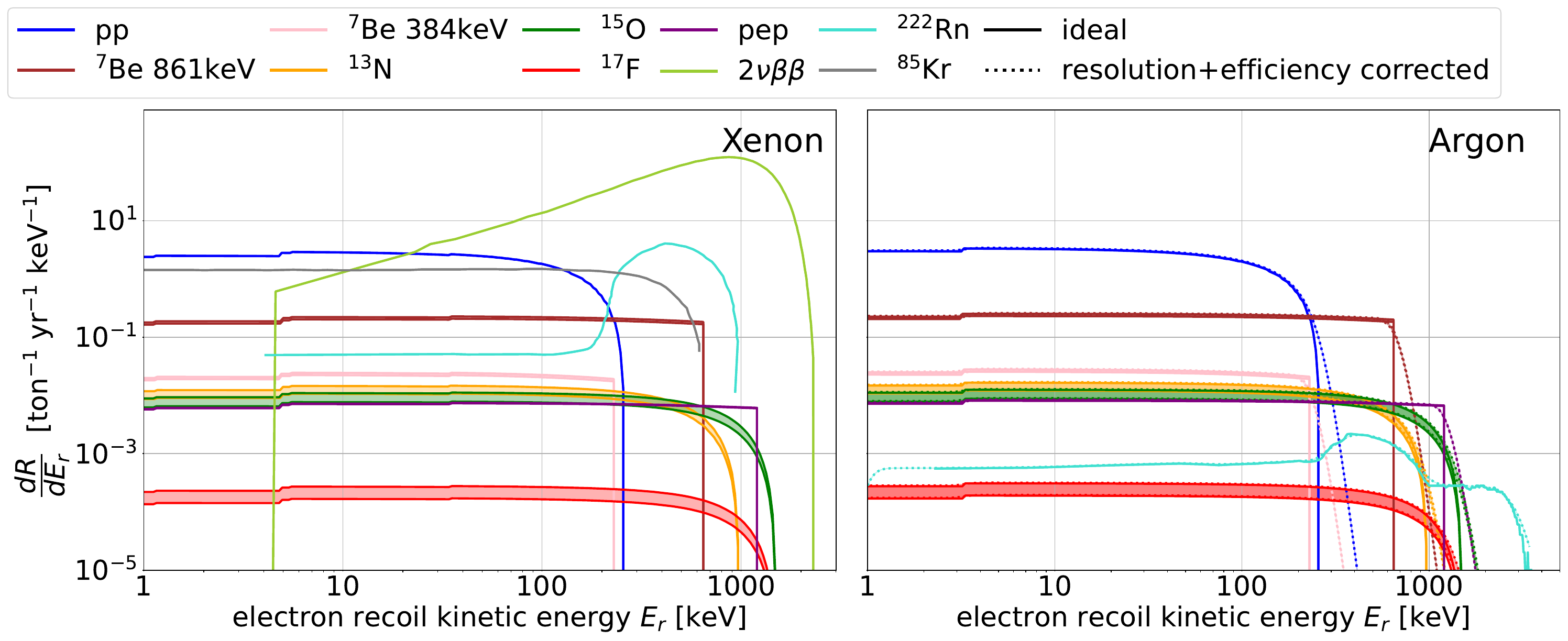}
\caption{ Neutrino-electron elastic scattering (ES) spectra for Xenon (left) and Argon (right) for solar and background components. Where discernible, the shaded regions reflect the difference between the high and low metallicity solar normalizations. The spectra with dashed curves for Argon are modified to account for energy smearing. For Xenon, efficiency and smearing are handled via the NEST simulations, as described in the text, and the only ideal spectra is shown here.
}
\label{fig:recoil_pdf_cdf_ER}
\end{figure*}

\begin{figure*}[!htbp]
\includegraphics[width = 0.95\textwidth ]{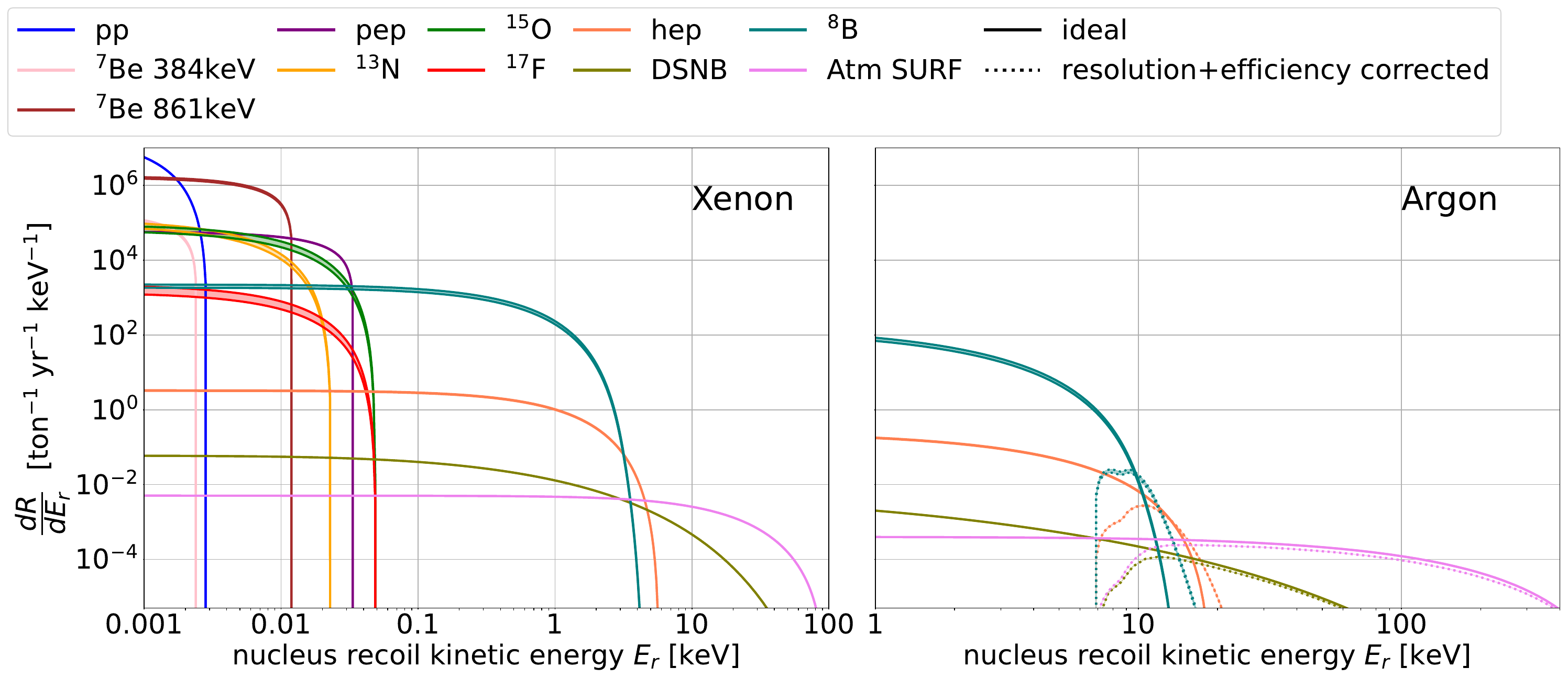}
\caption{Nuclear recoil spectra for Xenon (left) and Argon (right) via CE$\nu$NS. Shown are the components of the solar, atmospheric, and DSNB spectra. For the solar components, the shaded region reflects the difference between the high and low metallicity normalizations. The atmospheric spectra are shown for the SURF detector location. The spectra with dashed curves for Argon are modified to account for energy smearing and detector efficiency. For Xenon, efficiency and smearing are handled via the NEST simulations, as described in the text, and the only ideal spectra are shown here. }
\label{fig:recoil_pdf_cdf_NR}
\end{figure*}

 Figures~\ref{fig:recoil_pdf_cdf_ER} and   ~\ref{fig:recoil_pdf_cdf_NR} show the electron and nuclear recoil spectra, respectively, for the solar, atmospheric, and DSNB spectra. We show results for both for Xenon and Argon targets, which are the most likely target nuclei for large scale detectors with size $\sim 10-100$ ton. Physical processes that produce each flux component are shown in Table~\ref{tab:flux_bg_activity}. The nuclear recoil spectrum uses the neutral current coherent elastic neutrino-nucleus scattering (CE$\nu$NS) channel, and the electron recoil channel uses neutrino-electron elastic scattering (ES). The differential cross section and minimum neutrino energies are 
\begin{equation}
\begin{split}
    \frac{d\sigma (E_r, E_{\nu})}{dE_r} &=\left\{
    \begin{array}{cr}
     \frac{G_f^2 m_e}{2\pi}
 \left[(c_v + c_a)^2 + (c_v - c_a)^2 \left(1-\frac{E_r}{E_{\nu}}\right)^2 + ({c_a}^2 - {c_v}^2) \frac{m_e E_r}{E_{\nu}^2}\right] & \textrm{ES}\\
     \frac{G_f^2}{4\pi} Q^2 m_N  \left(1- \frac{m_N  E_r}{2E_{\nu}^2}\right) F^2(E_r) & \textrm{CE$\nu$NS}
    \end{array} \right.\\
    E_{\nu,min} &=\left\{
    \begin{array}{cr}
     \frac{1}{2}\left[E_r +\sqrt{E_r (E_r+2m_e)}\right]&\mbox{ES}\\
     \sqrt{\frac{m_N E_r}{2}}&\mbox{CE$\nu$NS}
    \end{array}\right.
\end{split}
\nonumber 
\end{equation}
where $E_r$ is the electron or nuclear recoil kinetic energy and $m_N$ is the mass of the target nucleus. Assuming pure Standard Model interactions, under ES, $g_v = 2 \sin^2\theta_w-1/2$, $g_a = -1/2$. We take the $\nu_e$ survival probability $P_{ee} = 0.553$~\cite{2017Chen}, which implies that $\sim 55.3\%$ of the $\nu_e$ flux remains when reaching the detector, which undergo both charged current and neutral current interactions. Then $\sim 45\%$ of $\nu_e$ flux oscillates to $\nu_{\mu,\tau}$ ($P_{\nu_{\mu,\tau}} = 1- P_{ee}$), which undergoes just neutral current interactions. So the total rate of ES is summing over these two contributions (Eqn~\ref{eqn: rate}). The difference in number of available Feynman diagrams appears in $\frac{d\sigma (E_r, E_{\nu})}{dE_r}$, where $\nu_e$ has $c_v = g_v + 1$, $c_a = g_a + 1$, and $\nu_{\mu,\tau}$ has $c_v = g_v$, $c_a = g_a$. On the other hand, CE$\nu$NS does not distinguish among flavors at tree level (Eqn~\ref{eqn: rate}). For CE$\nu$NS, \mbox{$Q = N - (1-4\sin^2\theta_w)Z$} and $\sin^2 \theta_w = 0.231$~\cite{ParticleDataGroup:2018ovx}, where $Z$ is the number of protons, $N$ is the number of neutrons, and the mean mass number $A=131.293$ for Xenon, and $A=39.948$ for Argon. We take $F(E_r)$ to be modelled by the Helm form factor~\citep{Lewin:1995rx}. 
Defining the neutrino flux as $d \phi /d E_\nu$, the rates of all flavors are then
\begin{equation}
\label{eqn: rate}
\begin{split}
\frac{dR_{\nu}(E_r)}{dE_r} 
&=\left\{
    \begin{array}{cr}
\frac{N_A}{A} \sum_{\nu_{\mu,\tau}, \nu_e} 
 \int_{E_{\nu,min}} dE_{\nu} \frac{d\phi}{dE_{\nu}}\frac{d\sigma (E_r, E_{\nu})}{dE_r} P_{\nu}&\mbox{ES}\\
 \frac{N_A}{A}
 \int_{E_{\nu,min}} dE_{\nu} \frac{d\phi}{dE_{\nu}}\frac{d\sigma (E_r, E_{\nu})}{dE_r}  & \textrm{CE$\nu$NS}
\end{array}\right.
\end{split}
\end{equation}
with \mbox{$N_A$/A} being number of nuclei per mass of the target nuclei. As applied to the solar neutrino flux, pp, $^7$Be, CNO and pep are detected through ES, where the electron neutrino survival probability $P_{ee}$ is required for calculating the rates. On the other hand $^8$B, atmospheric neutrinos, DSNB and hep are detected through CE$\nu$NS. Since we are only considering tree level interactions, there is no difference among for the interactions amongst the different $\nu_e$, $\nu_{\mu}$, $\nu_{\tau}$ flavors, so we do not need to consider the survival probability.

\par We use solar neutrino flux normalizations from GS98-SFII (high metallicity) and AGSS09-SFII (low metallicity)~\citep{Haxton:2012wfz}. The atmospheric fluxes used are the  average of the solar min and solar max flux calculated in~\cite{Zhuang:2021rsg} for different locations. DSNB event rate are obtained from~\cite{2009Strigari}. For the electron recoil component, we apply the electron binding energy correction which adds the step-like features at the lowest recoil energies~\citep{2017Chen}. Also shown in the electron channel are the projections for the $^{85}$Kr, $^{222}$Rn, and 2$\nu\beta\beta$ backgrounds~\cite{Franco:2015pha,2019Jayden}. There may be additional backgrounds to consider, for example low-energy beta emitters
produced in cosmic muon spallation, or other radiogenic backgrounds associated with the target medium~\cite{Aalbers:2022dzr}. Some of these backgrounds may vary with cosmic ray flux and therefore depth of detector, so they may introduce a location dependent effect in the backgrounds. Simulating these backgrounds are beyond the scope of our present analysis, which is mainly focused on the varying astrophysical backgrounds. 

\begin{table}[!htbp]
\caption{Physical process the produce the neutrino fluxes and backgrounds~\cite{Fukugita:2003en,1989Bahcall}
}
\begin{tabular}{c|c}
\hline
$\nu$ flux & physical process
\\

\hline
$pp$ & $p$ + $p$ $\rightarrow$ $d$ + $e^+$ + $\nu_e$ + $0.42 MeV$ \\
$pep$ & $p$ + $e^-$ + $p$ $\rightarrow$ $d$ + $\nu_e$ + $1.442 MeV$ (monoenergetic) \\
$^{7}Be$ & $^{7}Be$ + $e^-$ $\rightarrow$ $^{7}Li$ + $\nu_e$ + $0.862 MeV$ (monoenergetic) \\
$^8B$ & $^{7}Be$ $\rightarrow$ $^8B$ +$\gamma$ + $0.137 MeV$, $^8B$ $\rightarrow$ $^8Be^{*}$ + $e^+$ + $\nu_e$ + $15.04 MeV$\\
$hep$ & $^3He$ + $p$ $\rightarrow$ $^4He$ + $e^+$ + $\nu_e$ + $18.77 MeV$ \\
\multirow{3}{*}{$CNO$} & $^{13}N$ $\rightarrow$ $^{13}C$ + $e^+$ + $\nu_e$ \\
& $^{15}O$ $\rightarrow$ $^{13}N$ + $e^+$ + $\nu_e$ \\
& $^{17}F$ $\rightarrow$ $^{17}O$ + $e^+$ + $\nu_e$ \\
\multirow{2}{*}{${\rm Atm}$}  & $\pi^{+}$ $\rightarrow$ $\mu^{+}$ + $\nu_{\mu}$, $\mu^{+}$ $\rightarrow$ $e^{+}$ +  $\nu_{e}$ + $\overline{\nu}_{\mu}$\\
& $\pi^{-}$ $\rightarrow$ $\mu^{-}$ + $\overline{\nu}_{\mu}$, $\mu^{-}$ $\rightarrow$ $e^{-}$ +  $\overline{\nu}_{e}$ + $\nu_{\mu}$ \\
\multirow{2}{*}{2$\nu\beta\beta$} & $^{136}Xe$ $\rightarrow$ $e^-$ + $\overline{\nu}$ + $^{135}Cs$ \\
& $^{135}Cs$ $\rightarrow$ $e^-$ + $\overline{\nu}$ + $^{136}Ba$ \\
$^{85}Kr$ & $^{85}Kr$ $\rightarrow$ $^{85}Rb$ + $e^-$ + $\overline{\nu}$\\
$^{222}Rn$~\cite{Franco:2015pha}~\cite{2019Jayden} & $^{218}Po$ $\rightarrow$ $^{214}Pb$ + $\alpha$,  $^{214}Pb$ $\rightarrow$ $^{214}Bi$ + $e^-$ + $\overline{\nu}$\\
\hline
\end{tabular}
\label{tab:flux_bg_activity}
\end{table}

\section{NEST simulation and Detector Efficiency Models}
\label{sec:NEST}
\par To simulate the detection of the signal in Xenon time-projection chambers (TPC), we use the Noble Element Simulation Technique (NEST)~\citep{2011NEST} code. Neutrinos (or dark matter particles) interact with the liquid Xenon or Argon in the detector, producing a scintillation signal, $S1$, and ionization electrons, which then drift along the electric field to produce a signal, $S2$. The NEST code simulates the detection of events in the space of $S1$ and $S2$. For the NEST configuration, we choose all enhanced parameters (Table~\ref{tab:detector_file}), similar to previous studies ~\citep{2021Jayden,Tang:2023xub}, and adopt unit [phd] in our analysis. For the detector geometry, we take the radius to be $1300$ mm, and the $z$ position to range from $75.8$ mm to $1536.5$ mm. 

\par
\begin{figure*}[!htbp]
\includegraphics[width = 0.45\textwidth ]{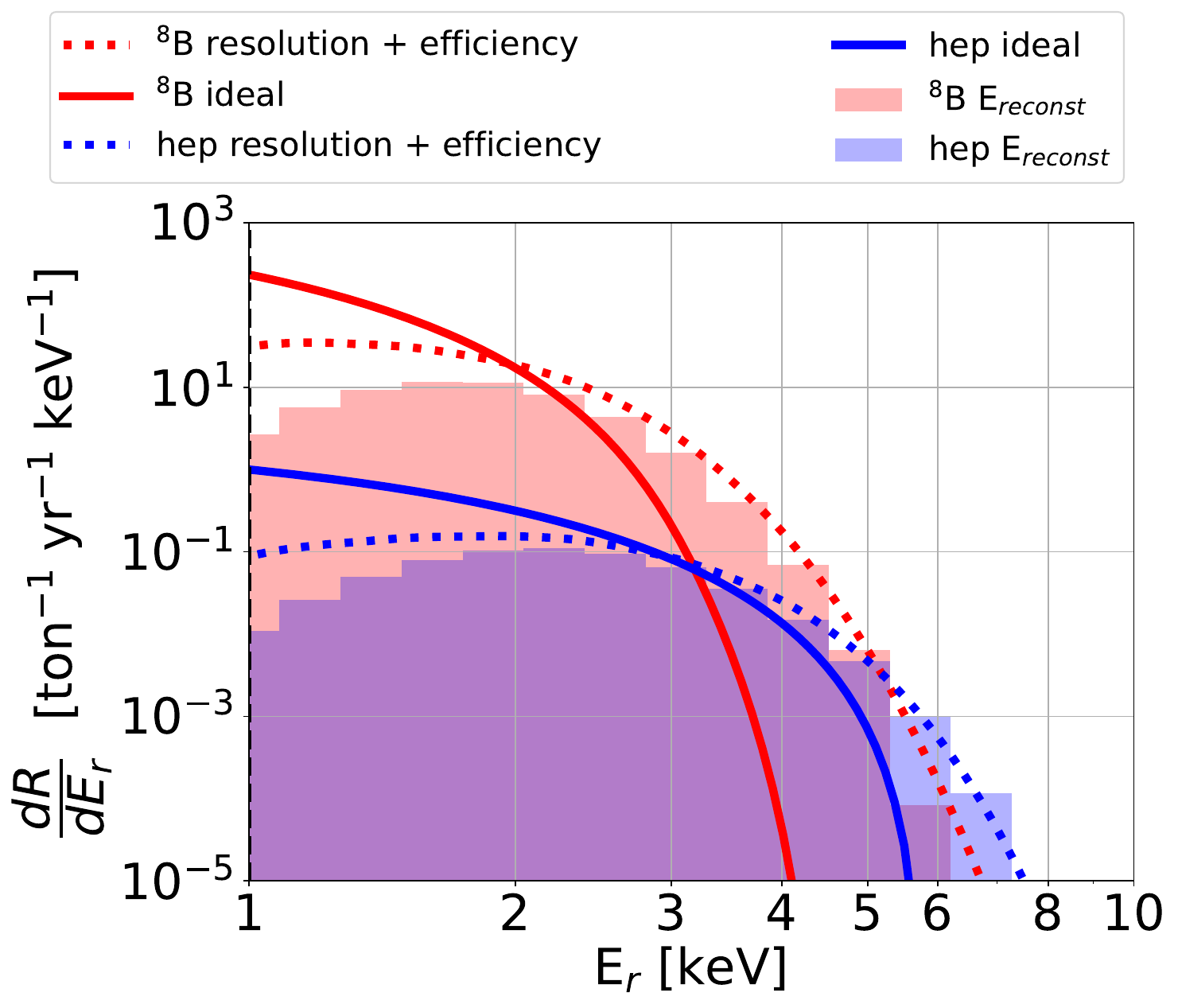}
\caption{Comparison between analytical model and reconstructed event rate from NEST. }
\label{fig:compare}
\end{figure*}

 For comparison to the analysis in $S1/S2$ space, we will perform an analysis directly in electron and nuclear recoil space.  With the efficiency modeled as a function of recoil energy, the modified event rate is 
\begin{equation}
    \frac{dR}{dE_{r}} = \epsilon (E_{r}) \int dE_{r}^{'} \frac{dR(E_{r}^{'})}{dE_{r}^{'}}
     \frac{1}{\sqrt{2\pi \sigma^{2} (E_{r}^{'})} } e^{-\frac{(E_{r} - E_{r}^{'})^2}{ 2\sigma ^{2}  (E_{r}^{'})}}, 
     \nonumber 
\end{equation}
where $E_{r}^{'}$ is the true recoil energy, $E_{r}$ is the detected recoil energy, $\sigma(E_{r}^{'})$ is the resolution at the true recoil energy, and $\epsilon (E_{r})$ is the detector efficiency. The comparison between the modified event rate obtained through the``resolution + efficiency" analytic model and that through the NEST simulation is shown in Figure~\ref{fig:compare}. In the analytic model, the resolution is \mbox{$\sigma(E_{r}) =\left (0.31\sqrt{\frac{E_{r}}{\textrm{keV}}} + 0.0035\frac{E_{r}}{\textrm{keV}} \right) \, {\rm keV}$}~\cite{2019Jayden} and the efficiency is the probability of binned events with valid S1 and S2 signals out of uniformly simulated binned events. 

\par For Xenon and Argon, we consider four models  defined by how the detector efficiency and energy resolution are modeled. These models are defined as follows: 
\begin{enumerate}
    \item Ideal Xenon: Characterized by perfect efficiency and energy resolution. 
    \item Xenon $S1/S2$: Full analysis in $S1/S2$ space for Xenon, including efficiency and energy resolution corrections via the NEST simulations. Due to the minimum work function in the NEST setup (13.469 eV) to excite the atom, nuclear recoils below this energy do not generate $S1/S2$ signals. So the solar CE$\nu$NS components with neutrino energies at this scale do not contribute to the rate in this scenario. 
    \item Ideal Argon: Characterized by perfect efficiency and energy resolution.  
    \item Argon resolution + efficiency: Characterized by an energy resolution of 
    \mbox{$\sigma(E_{r}) = 0.1E_{r}$}. The nuclear recoil efficiency is taken from DarkSide-50~\citep{2016PhRvD..93h1101A}. For electron recoils, we simply assume a threshold energy of $E_{th} = 1$ keV.   
\end{enumerate}

\section{Statistical Methods \label{sec:statistical}}
\par In this section, we establish our statistical methods for the detection of solar and atmospheric neutrinos. For related previous analysis, see Refs.~\citep{OHare:2021utq,2014Ruppin} based on the statistical methodology in Ref.~\citep{2011Cowan}. As described above, there are two components to our analysis. The first involves an Argon-based and idealized Xenon-based analysis in the nuclear and electron recoil energy ($E_r$) space, and the second involves a Xenon-based analysis in the $S1/S2$ space. 

\par We start with construction of the function applicable to both $E_r$ and $S1/S2$ space. Let $\Omega$ represent the entire domain of interest, which can exist in one-dimensional, two-dimensional, or ultimately event higher-dimensional space, depending on the observable. For instance, in the analysis in recoil energy space in Argon and Xenon, $\Omega$ is one-dimensional, and in the analysis of Xenon in $S1/S2$ space, $\Omega$ is two-dimensional. To detect the presence of a given neutrino flux component, multiple non-overlapping subdomains $\Omega_i$, $i=1,2,\cdots, M$ are selected from $\Omega$. Let the detector have a volume of $\D$ and a run-time of $\T$. In the $i$-th subdomain $\Omega_i$, let $\eta_i^\kappa$ be the mean event rate for a flux component, measured in units ton$^{-1}$ yr$^{-1}$, so that the expected number of events for all components is 
\begin{equation}
\mu_{i} = \D\T \sum_{\kappa}\eta_i^\kappa. 
\label{eq:events}
\end{equation}
Defining the observed number of events as $n_i$ at $\Omega_i$, the likelihood for the observed number of events over $M$ subdomains is
\begin{equation}
               \mathcal{L} = \prod_{i=1}^{M} \frac{\exp(-\mu_{i} )\mu_{i} ^{n_{i}} }{n_i!},
   \nonumber
\end{equation} 
where the observations across the subdomains are assumed to be independent.

\subsection{Analysis in E$_{r}$ and S1$/$S2 space }

\par We start with an analysis directly in $E_r$ space for the detected electron and nucleus. We use this $E_r$ space for the idealized Xenon, idealized Argon, and Argon resolution+efficiency analysis methods. In this case, the domain is one-dimensional, and the $\imath^{th}$ subdomain $\Omega_i$ corresponds to an energy range in the one-dimensional $E_r$ space. The event rate in the $\imath^{th}$ subdomain from the $\kappa^{th}$ flux component is 
\begin{equation}
\eta_i^\kappa = \int dE_r \frac{dR_{i}^\kappa}{dE_r} 
\nonumber
\end{equation}
where the integral is over the energy range covered by the $\imath ^{th}$ subdomain. The expected number of events is then derived using Equation~\ref{eq:events}.  

\par We then move on to a Xenon analysis in $S1/S2$ space. In this case, the domain $\Omega$ is two-dimensional, and has total $N_{S1}$ S1 bins and $N_{S2}$ S2 bins (in log$_{10}$ space). We construct a likelihood function in terms of the detector-related $S1/S2$ variables. For the $\kappa^{th}$ flux component, we can schematically represent the event rate in the $i$-th subdomain $\Omega_i$ in the two-dimensional space as 
\begin{equation}
    \eta_{i}^\kappa = \int d S1 d S2 \frac{d^2 R_{i}^\kappa}{d S1 d S2} {\rm pdf} (S1,S2 |E_r).  
    \label{eq:dnds1ds2} 
\end{equation}
In Equation~\ref{eq:dnds1ds2},  
 ${\rm pdf} (S1,S2 |E_r)$ is the probability density function for $S1$ and $S2$ for a given $E_r$.  The pdf converts a recoil energy spectrum to $S1/S2$ space after the interaction process in the detector, and is calculated from the NEST simulations described above. To determine the pdf for each component, $10^7$ energy depositions are simulated and converted to $S1/S2$ signals. Since each component has its own recoil spectrum, this leads to a unique event rate in the subdomain $\Omega_i$ of $S1/S2$ space.

\subsection{Test of significance}
\label{sec:TestSig}

\par Our goal will be to detect the solar, atmospheric, and DSNB flux components. To achieve this, we define components of the flux as the background to be contained in the null hypothesis, $H_0$. We then add a flux component $\eta_{i}^\gamma$ on top of the background in the null to define the alternative hypothesis, $H_1$. As an example, if we are attempting to detect hep neutrinos via nuclear recoils, the null hypothesis $H_0$ includes $^{8}$B, DSNB, and atmospheric neutrinos. The alternative hypothesis $H_1$ includes $^{8}$B, DSNB, atmospheric, and hep neutrinos.

\par Now define the total number of events over all subdomains of interest or being selected as $N=\sum_{i}n_{i}$; here index $i$ is used to denote different sub-regions. 
Then the expectation of the random variable $N$ is 
\begin{equation}\label{test:simple}
    E(N) = \left\{
    \begin{array}{lr}
     {\cal D}{\cal T} \sum_{i} \sum_{\kappa} \eta_{i}^\kappa  &\mbox{under }H_0,\\
      {\cal D}{\cal T}  \sum_{i} (\sum_{\kappa} \eta_{i}^\kappa +\eta_{i}^\gamma) &\mbox{under }H_1. 
    \end{array}\right.
\end{equation}
The goal of the following exercise is to find the exposure $\D\T$ that ensures  $H_0$  is rejected when $H_1$ holds with a desired probability while the test's significance level is set to $\alpha$. 
Moreover, here we assume that $\eta^\kappa_i$
and $\eta^\gamma_i$ are known for all $i$ and $\kappa$. The index $\kappa$ is used to to denote different background components.

\par The level or the significance level $\alpha$ denotes the probability of Type-I error, and Type-I error signifies rejection of $H_0$ when $H_0$ holds. In the context of this problem, the Type-I error signifies the situation where based on the data and statistical test, we declare the presence of a particular neutrino signal $\eta^\gamma_i$, but in reality there were no such signals other than the background noise. On the other hand, the probability of a type-II error denoted by $\beta$ signifies the probability of failing to reject $H_0$ when $H_1$ holds. In the context of this problem, Type-II error signifies the situation where based on the data and statistical test, we failed to find the presence of a neutrino signal $\eta^\gamma_i$, but in reality, it was present along with the background noise. The power of the test is one minus the probability of Type-II error (i.e., power$=1-\beta$). The probabilities of a Type-I error and a Type-II error are in inverse relation. 

Given that $\eta_{i}^\gamma \geq 0$ for all 
$i$, the most powerful test for testing the simple null against the simple alternative as given in Equation~\ref{test:simple} at level $\alpha$ is reject $H_0$ if $N>N_\alpha=N_\alpha({\cal D}{\cal T})$, the smallest integer of the set 
\begin{equation}\nonumber\label{eq:foralpha} 
\left[\scriptr:  \pr \{N> \scriptr \mbox{ where }N\sim {\rm Poisson}({\cal D}{\cal T} \sum_{i} \sum_{\kappa} \eta_{i}^\kappa) \}\leq \alpha\right]. 
\end{equation}
This form of the test is obtained by using the Neyman-Pearson lemma and the monotone likelihood ratio property of the Poisson distribution \citep[p.~388, 391]{CB}. Note that there is no closed form analytical expression for $N_\alpha$ in terms of ${\cal D}{\cal T} \sum_{i} \sum_{\kappa} \eta_{i}^\kappa$, so it must be computed numerically. However, there is no need to do a simulation to compute $N_\alpha$. Moreover, this test is uniformly the most powerful test for testing the null $H_0: E(N)= \D\T\sum_i\sum_\kappa \eta^\kappa_i$ against the alternative $H_1: E(N)> \D\T\sum_i\sum_\kappa \eta^\kappa_i$. For a minimum power of $1-\beta$ for the alternative $E(N)= \D\T\sum_i(\sum_\kappa \eta^\kappa_i+\eta^\gamma_i)$, the desired ${\cal D}{\cal T}$ is 
\bse 
{\inf}\left[ {\cal D}{\cal T}:   \pr \left\{N> N_{\alpha} \mbox{ where }N\sim {\rm Poisson}\left({\cal D}{\cal T}  \sum_{i} (\sum_{\kappa} \eta_{i}^\kappa +\eta_{i}^\gamma)\right) \right\}
\geq  (1-\beta) 
\right].
\ese 
To find this exposure $\D\T$, we first fix $\alpha$ and $(1-\beta)$, then for a grid of values of $\D\T$, find $N_\alpha$. For each $\D\T$ and the corresponding $N_\alpha$, we compute 
$$
\pr \left\{N> N_{\alpha} \mbox{ where }N\sim {\rm Poisson}\left(\D\T \sum_{i} (\sum_{\kappa} \eta_{i}^\kappa +\eta_{i}^\gamma)\right) \right\},
$$
and check if the above probability is at least $(1-\beta)$. The infimum of the set of $\D\T$ that yields the above probability to be at least $(1-\beta)$ is the desired exposure.

\subsection{Choice of the subdomain} 

\par Given the nature of the distribution of the events in the subdomains, the null and alternative hypotheses may be difficult to distinguish, especially when a large portion or almost all of the domain with a very small signal is on top of a relatively large background. This is especially true for the CE$\nu$NS channel since the atmospheric, hep and DSNB event rates are low. It is less of a concern for the ES channel due to their relatively large event rates. For both cases, we wish to choose a set of subdomains that maximize the ability of the test to differentiate $H_0$ and $H_1$. With this motivation, we now discuss choices of subdomain for each of the analysis methods. One recommended principle for selecting subdomains is to enhance the signal-to-background noise ratio within the chosen subdomains when compared to using the entire domain without any selection while keeping sufficient exposure. In order to characterize the background components included in the analysis, we introduce the following notation
\begin{enumerate}
    \item \mbox{$\kappa_{all, S1/S2}$ = (pp, $^{7}$Be 861, $^{7}$Be 384, pep, CNO, $2\nu\beta\beta(f_{2\nu\beta\beta})$, $^{85}$Kr, $^{222}$Rn, $^{8}$B, hep, DSNB, Atm)}
    \item \mbox{$\kappa_{all, ER}$ = ($^{7}$Be 861, CNO, pep)}
    \item \mbox{$\kappa_{all, NR}$ = (Atm, $^{8}$B, hep, DSNB)}
\end{enumerate}

\subsubsection{Choice of two-dimensional subdomain in S1/S2 space for ES Xenon}\label{sec:S1/S2_ERbinning}

\par For the detection of the CNO and pep components via elastic scattering in Xenon, we divide into 30 equally spaced $S1$ bins within the range $[2-12000]$, and 30 equally spaced log$_{10}$S2 bins in the range $[2-6.9]$. We do not select subdomain since the event rate of CNO and pep are $\sim O(10)$. 
The null hypothesis includes flux components $\eta^\kappa$, when considering the detection of the CNO and pep fluxes in the electron scattering channel (\mbox{$\gamma$ = CNO or pep}). This considers Xenon as the nuclear target and a full analysis in the S1/S2 space (Table~\ref{tab:ER_bg}). 

\begin{table}[!htbp]
\caption{Flux components $\eta^\kappa$ that are included in the null hypothesis, listed in the $\kappa$ column, when considering the detection of the CNO and pep fluxes in the electron scattering channel. This considers Xenon as the nuclear target and a full analysis in the S1/S2 space. The two rows of the second column indicate whether the background includes only $2\nu\beta\beta$ or all of $2\nu\beta\beta$, $^{85}$Kr and $^{222}$Rn. The ``-" sign in the third column indicates the components that have been removed from $\kappa_{all, S1/S2}$. 
}
\begin{tabular}{c|c|c}
\hline
$\gamma$ & Detector backgrounds & Background components ($\kappa$)
\\
\hline
\multirow{2}{*}{CNO} & $2\nu\beta\beta$ &$\kappa_{all, S1/S2}$ - $\gamma$ - $^{85}$Kr - $^{222}$Rn \\
& all  & $\kappa_{all, S1/S2}$ - $\gamma$\\
\cline{1-3}
\multirow{2}{*}{pep} & $2\nu\beta\beta$  &$\kappa_{all, S1/S2}$ - $\gamma$ - $^{85}$Kr - $^{222}$Rn\\
& all  & $\kappa_{all, S1/S2}$ - $\gamma$\\
\hline
\end{tabular}
\label{tab:ER_bg}
\end{table}

\subsubsection{Choice of two-dimensional subdomain in S1/S2 space for CE$\nu$NS Xenon}
\label{sec:S1/S2_NRbinning}

\par For the detection of atmospheric, hep and DSNB components through the nuclear recoil channel, we take additional steps to define the optimal set of subdomains in $S1/S2$ space. This is because the backgrounds contained in the null hypothesis can be significant, and care must be taken to identify the signal. 

\par We start by dividing $S1/S2$ space into a large number of small regions, and then combine the small regions into a subdomain. More specifically, for a fixed $S1$, we combine the consecutive log$_{10}$S2 regions in which the signal is greater than the background into one rectangular log$_{10} S2$ subdomain. When scanning over the entire $S1/S2$ space, we keep only the subdomains with similar event rates so that the total rate summed over all subdomains is nearly uniform. We do not include bins with a very small event rate, corresponding to a threshold of signal $>$ 10$^{-4}$ ton$^{-1}$ yr$^{-1}$. 

\par
\begin{figure*}[!htbp]
\includegraphics[width = 0.9\textwidth ]{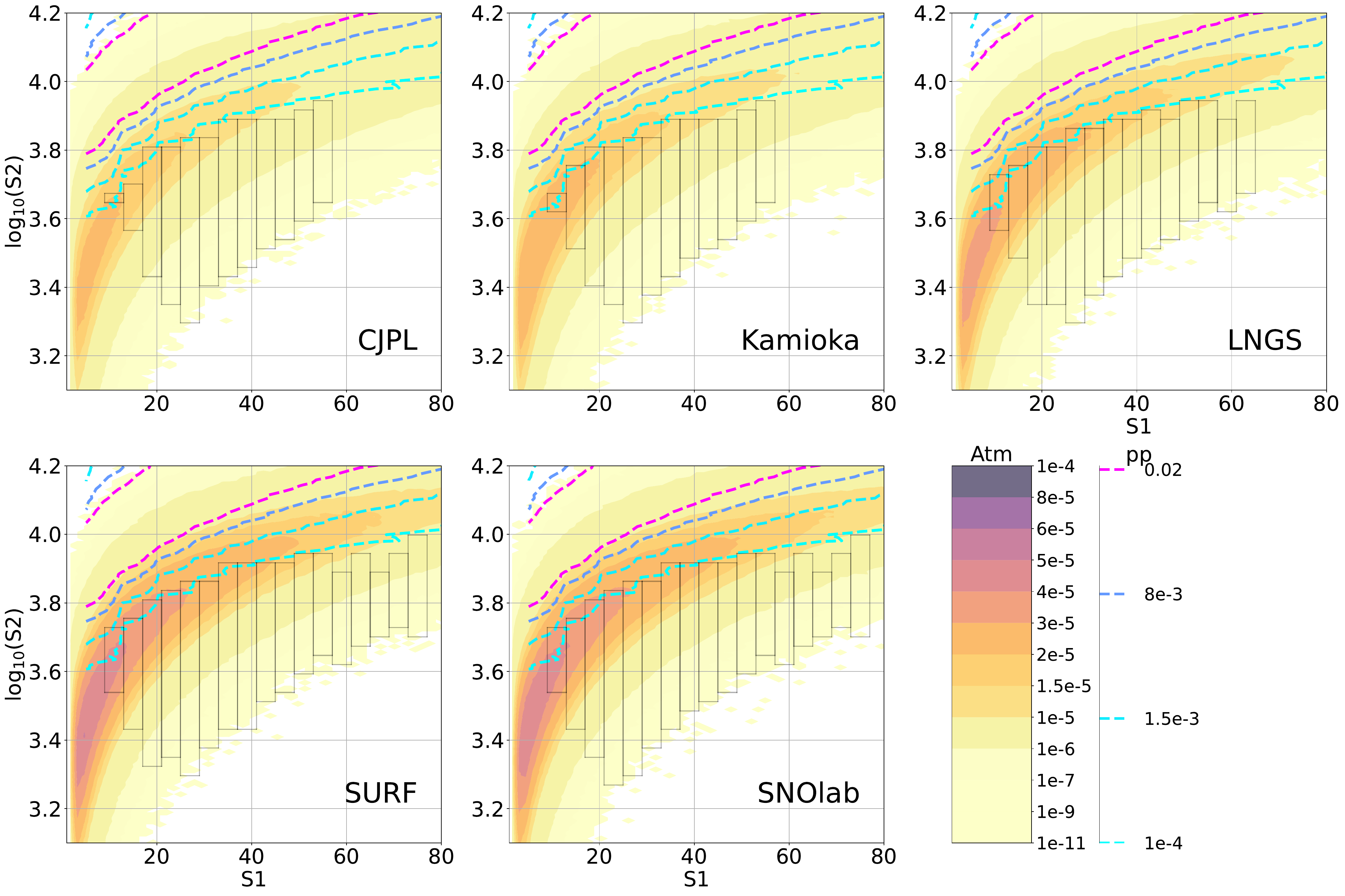}
\caption{Heat map of the atmospheric neutrino event rate from CE$\nu$NS, and contours of constant pp event rate from elastic scattering (dashed contours) in the S1/S2 plane. The event rates are in units of ton$^{-1}$ yr$^{-1}$. The panels are for the five different detector locations that we consider. Black rectangles indicate the bins used in the analysis for each of the locations.}
\label{fig:atmNu_method2_region}
\end{figure*}

 As an example of this procedure, consider the detection of the atmospheric neutrino signal. In this case, the null hypothesis is given by the background components in Table~\ref{tab:optimized_NR_bg}. We simulate 10$^7$ $S1/S2$ events using NEST for each CE$\nu$NS and ES component. We then divide the events into many equally spaced small S1/S2 grids, where number of the grids and the S1/S2 range are specifically shown in Table~\ref{tab:optimized_NR_bg}. This example of atmospheric neutrinos highlights the motivation for starting with small regions in log$_{10}$S2 space, which in this case is to avoid a significant contamination of events from leakage due to ES of pp solar neutrinos. The resulting set of subdomains for atmospheric signal are shown in Figure~\ref{fig:atmNu_method2_region} at each detector location. 

\begin{table}[!htbp]
\caption{Flux components $\eta^\kappa$ that are included in the null hypothesis, listed in the $\kappa$ column, when considering the detection of the atmospheric, hep and DSNB fluxes in the CE$\nu$NS channel at detector locations CJPL, Kamioka, LNGS, SURF, SNOlab. The ``-" sign in the second column indicates the components that have been removed from $\kappa_{all, S1/S2}$. Xenon is the nuclear target and the analysis is in $S1/S2$ space. The last two columns indicate the $S1/S2$ range used to generate the grid and the number of grids. }
\begin{tabular}{c|c|cc|cc}
\hline
$\gamma$ &  Background components ($\kappa$) & \# equally spaced grid  & S1 range & \# equally spaced grid & log$_{10}$S2 range\\
\hline
Atm & \multirow{3}{*}{$\kappa_{all, S1/S2}$ - $\gamma$} 
& 100  & $[1-401]$ & 100 & $[2-4.7]$ \\

hep  & 
& 10 & $[1-25]$ & 10 & $[1.9-4]$\\

DSNB & 
& 90 & $[1-361]$ & 80 & $[1.9-4.6]$\\
\hline
\end{tabular}
\label{tab:optimized_NR_bg}
\end{table}

\begin{figure}[!htbp]
\includegraphics[width = 0.45\textwidth ]{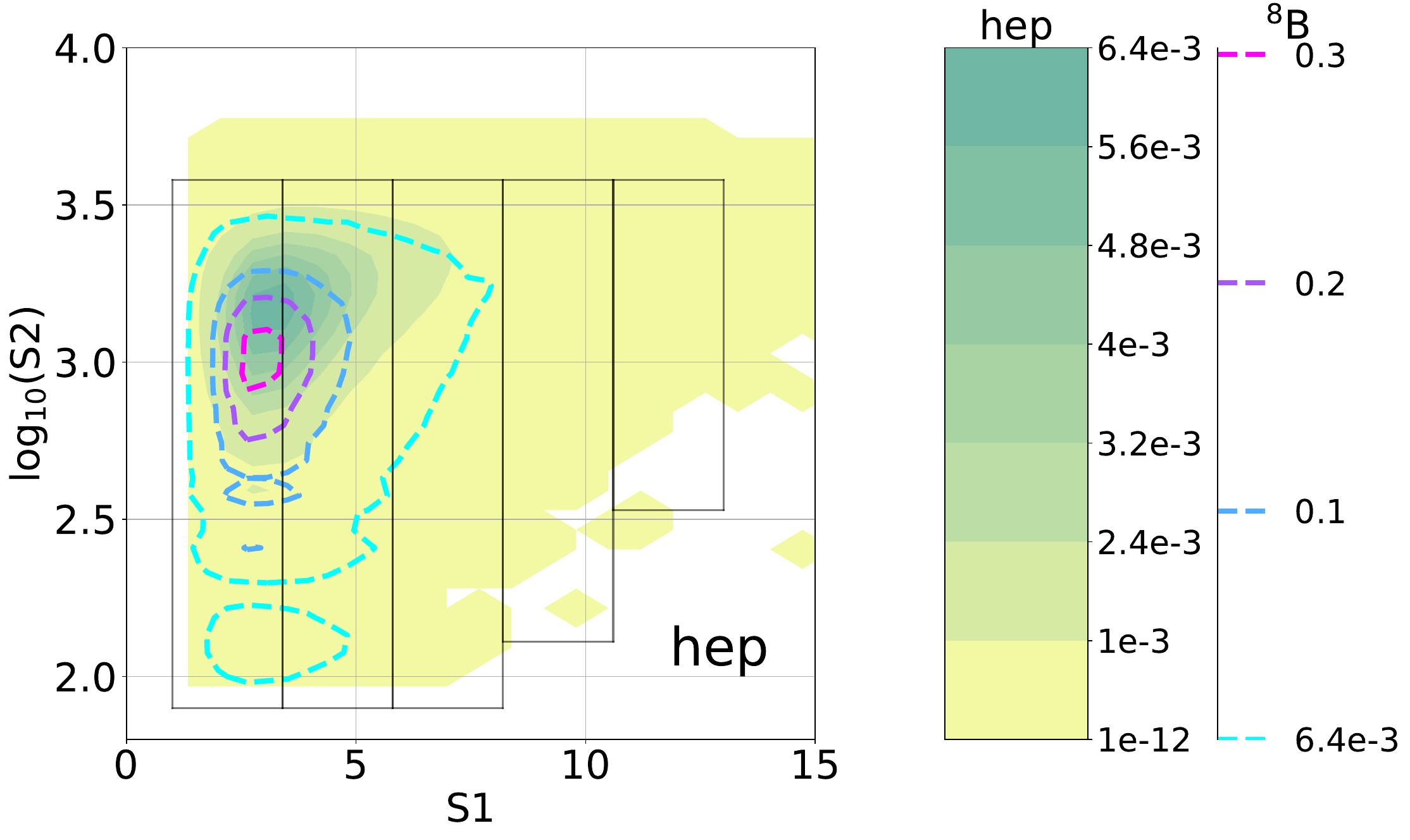}
\includegraphics[width = 0.45\textwidth ]{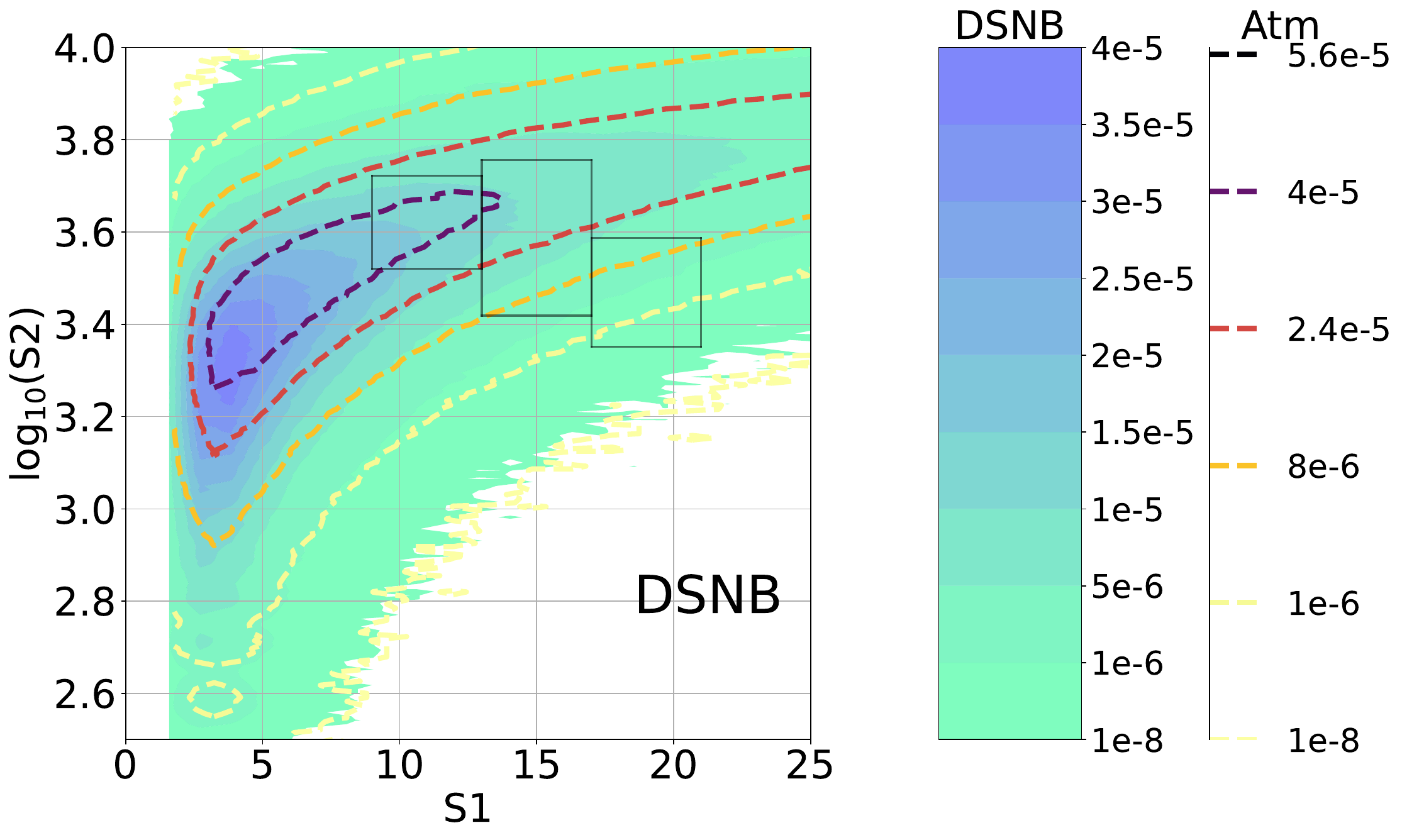}
\caption{Left: Heat map of the hep event rate and contours of constant $^8$B event rate (dashed). Right: Heat map of the DSNB event rate and contours of constant atmospheric event rate (dashed). The event rates are in units of ton$^{-1}$ yr$^{-1}$. The atmospheric event rate is for the SURF location. Black rectangles indicate the bins used in the analysis
}
\label{fig:hepdsnb_method2_region}
\end{figure} 

\par For the hep and DNSB components, we slightly adjust our method for determining the set of subdomains. This is because to detect the DSNB (hep) signal, the atmospheric background ($^8$B) is nearly entirely overlapping the regions where these signals are located, and the event rate of the DSNB (hep) is smaller than the atmospheric ($^8$B) background. There is no region in the $S1/S2$ space over which these components are larger than the background components. Therefore, in choosing the set of subdomains to search for the DSNB signal, we follow the same steps as above, except that we exclude the atmospheric signal. Note that in the full analysis for the DSNB, we do include the atmospheric component, we only exclude it to define the subdomains. 

\par For hep, we follow a similar strategy as for the DSNB, in this case excluding the more dominant $^{8}$B signal over that region. Then for the full analysis, we include the $^{8}$B component. The full background models considered for DSNB and hep are shown in Table~\ref{tab:optimized_NR_bg}. The results for the subdomains are shown in Figure~\ref{fig:hepdsnb_method2_region}. 

\subsubsection{Choice of one-dimensional subdomain in recoil energy space for ES and CE$\nu$NS}\label{sec:Er_ERbinning}

\par For the analyses in recoil energy space, for ES, the background components used for the detection of CNO and pep neutrinos are shown in Table~\ref{tab:Er_ER_bg}. This table may be compared to $\kappa$ in~\ref{sec:S1/S2_ERbinning} which shows the background components used to detect CNO and pep neutrinos in the $S1/S2$ Xe analysis. 

\par
\begin{figure}[!htbp]
\includegraphics[width = 0.9\textwidth ]{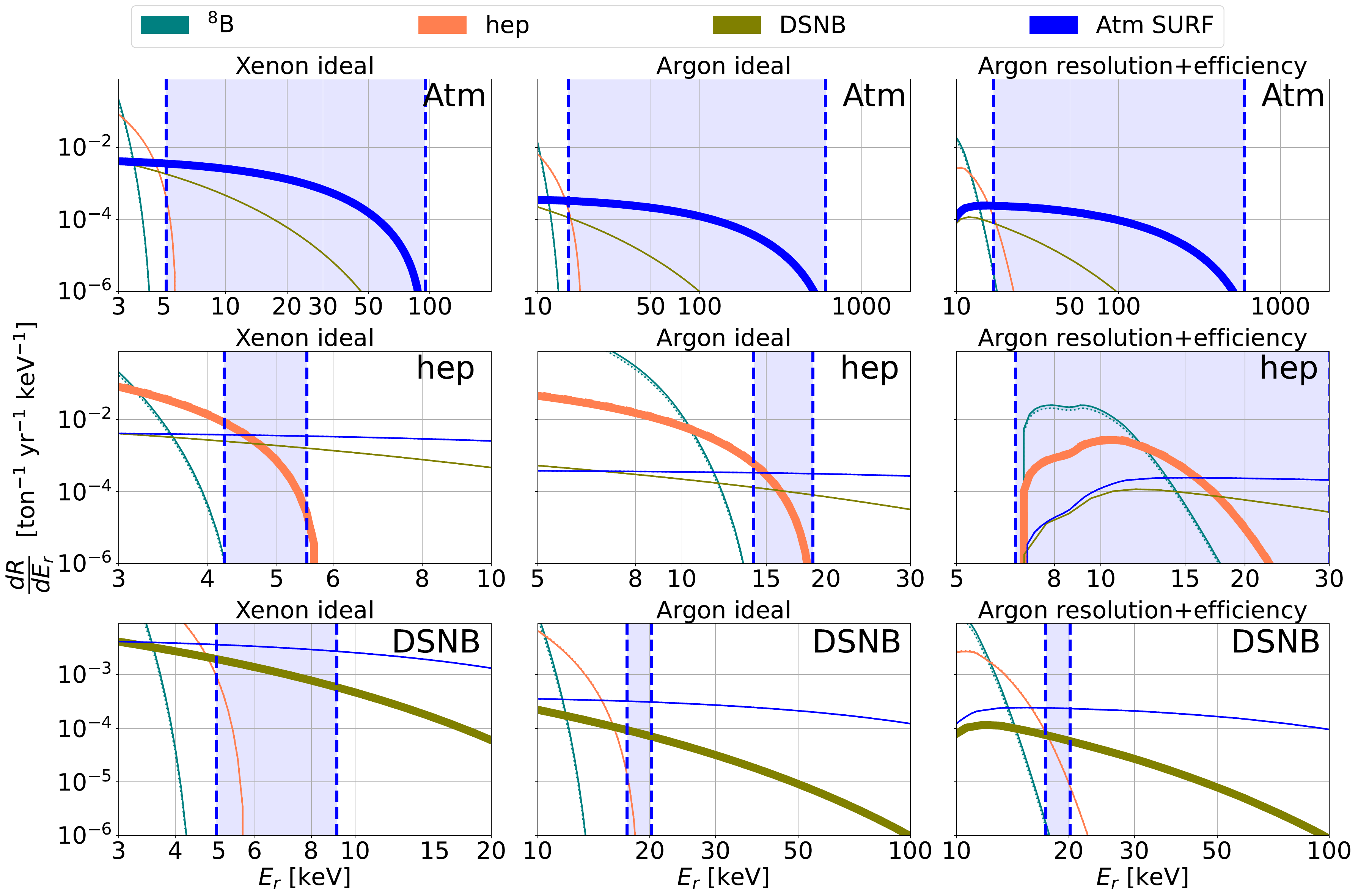}
\caption{Event rate spectrum and energy bin ranges used for the analysis for atmospheric (top row), hep (middle row), and DSNB (bottom row) versus recoil energy. The atmospheric neutrino spectrum is for the SURF location. The first column is for ideal Xenon, the second column is for ideal Argon, and the third column is for Argon resolution and efficiency correction. Blue shaded regions show the binning regions in recoil energy space according to the selection criteria outlined in the text and Table~\ref{tab:Er_NR_bg}, which are determined based on the simulations. 
}
\label{fig:NR_Er_region}
\end{figure}

 For CE$\nu$NS, similar to the $S1/S2$ analysis described above, the event rate is low in $E_r$ space. Upon optimizing the $E_r$ range as above, Figure~\ref{fig:NR_Er_region} shows the $E_r$ range used for the SURF detector location. Similarly, the criteria for selection of the $E_r$ ranges for all detector locations are shown in Table~\ref{tab:Er_NR_bg}. The criteria for selecting subdomain is obtained from testing several choices to select subdomain and find the criteria which gives the minimum exposure. 

\begin{table}[!htbp]
\caption{Flux components that are included in the null hypothesis, listed in the $\kappa$ column, when considering the detection of CNO and pep fluxes in the electron scattering channel, with the observable being the electron recoil energy. The ``-" sign in the last column indicates the components that have been removed from $\kappa_{all, ER}$, and ``+" sign indicates the components that have been added to $\kappa_{all, ER}$. The second column gives the nuclear target, Xenon or Argon. The third column indicates whether efficiency and resolution are included. The energy threshold is the end point of pp, which removes two background components, pp and $^{7}$Be 384, so we use $\kappa_{all, ER}$.
}
\begin{tabular}{c|c|c|c|c}
\hline
$\gamma$ & Nucleon & Analysis method & $E_r$ threshold [keV] & Background components ($\kappa$)\\
\hline
\multirow{4}{*}{CNO} & Xe & ideal & 291  &\multirow{2}{*}{$\kappa_{all, ER}$-$\gamma$}\\
\cline{2-4}
&\multirow{2}{*}{Ar} & ideal & 284 & 
\\
\cline{3-5}
&& resolution+efficiency& 414 &$\kappa_{all, ER}$ - $\gamma$ + $^{222}$Rn\\
\hline
\multirow{4}{*}{pep} & Xe & ideal & 291  &\multirow{2}{*}{$\kappa_{all, ER}$-$\gamma$}\\

\cline{2-4}
& \multirow{2}{*}{Ar}  & ideal& 284 & \\
\cline{3-5}
& & resolution+efficiency& 414 & $\kappa_{all, ER}$-$\gamma$ + $^{222}$Rn\\
\hline

\end{tabular}
\label{tab:Er_ER_bg}
\end{table}

\begin{table}[!htbp]
\caption{Flux components that are included in the null hypothesis, listed in the $\kappa$ column, when considering the detection of the atmospheric, hep and DSNB fluxes in the CE$\nu$NS channel in nuclear recoil space. The ``-" sign in the second column indicates the components that have been removed from $\kappa_{all, NR}$. The third column gives the nuclear target, Xenon or Argon. The fourth column indicates whether efficiency and resolution are included in the electron recoil energy space. The last column indicates how the energy range is selected. 
We use \mbox{$\kappa_{all, NR}$} as denoted before.
}
\begin{tabular}{c|c|c|c|c}
\hline
$\gamma$ & $\kappa$ & Nucleon & Analysis Method & Condition for the subdomain selection \\
\hline
\multirow{3}{*}{Atm} &\multirow{9}{*}{$\kappa_{all, NR}$ - $\gamma$}  & Xe  & ideal & \multirow{3}{*}
{$\eta_{i}^\gamma$ $>$ $\sum_{\kappa} \eta_{i}^\kappa $} \\

\cline{3-4}
&&  \multirow{2}{*}{Ar} & ideal & \\
\cline{4-4}
&&& resolution+efficiency &\\

\cline{1-1} \cline{3-5}

\multirow{3}{*}{hep} &  & Xe  &  ideal &  \multirow{2}{*}{ remove all $^{8}$B from pdf  simulations}\\
\cline{3-4}
&& \multirow{2}{*}{Ar} & ideal &   \\

 \cline{4-5}
&&& resolution+efficiency & keep all hep\\

\cline{1-1} \cline{3-5}

\multirow{3}{*}{DSNB } & & Xe  & ideal   & \multirow{3}{*}{$\eta_{i}^\gamma$ $>$ $\sum_{\kappa = hep, ^8B} \eta_{i}^\kappa $ and $\eta_{i}^\gamma$ $>$ $1/4 \eta_{i}^{atm}$ 
}\\

\cline{3-4}
&&\multirow{2}{*}{Ar} & ideal  \\

\cline{4-4}
&&& resolution+efficiency  & \\
 
\hline
\end{tabular}

\label{tab:Er_NR_bg}
\end{table}

\section{Results \label{sec:results}}
\par
\begin{figure*}[!htbp]
\includegraphics[width = 0.9\textwidth ]{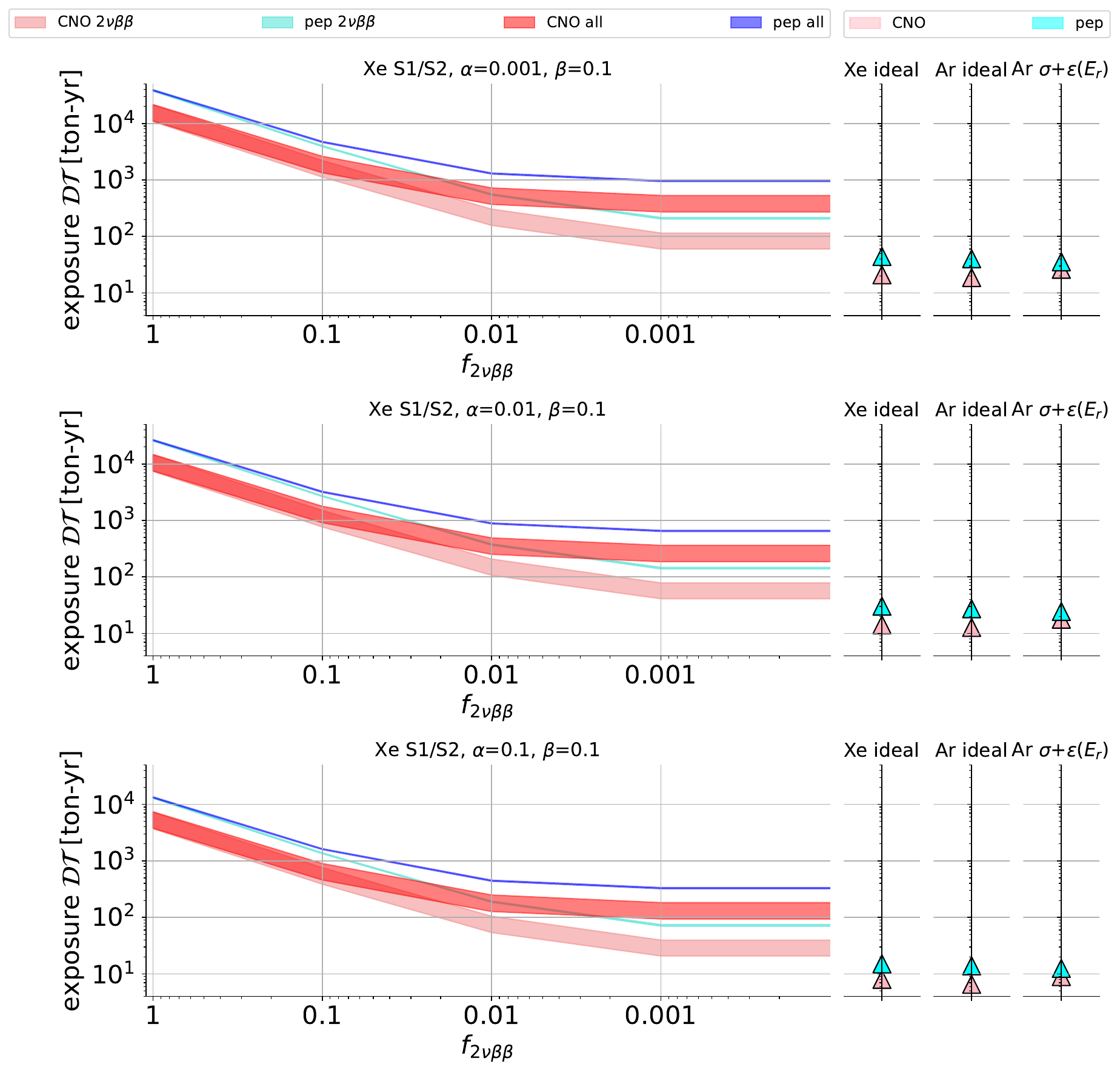}
\caption{The three panels with the curves show the prospects for detection of the CNO and pep fluxes in the elastic scattering channel for a given exposure as a function of the $2\nu\beta\beta$ fraction, $f_{2\nu\beta\beta}$, for Xenon, in the S1/S2 space. The bands are for high and low metallicities. The three different rows are for different values of Type I error rate $\alpha$ in the text. The three figures stacked on the right show the exposures for the ideal Xenon, Argon, and Argon resolution and efficiency models, for high metallicity. 
}
\label{fig:method2_DTvsALPHA_ER}
\end{figure*}

 We start by presenting the results for the analysis in the electron recoil channel, focusing on the CNO and pep neutrino fluxes. Figure~\ref{fig:method2_DTvsALPHA_ER} shows the detector exposure which satisfies $\alpha = 0.001, 0.01, 0.1$ and $\beta=0.1$, as a function of $f_{2\nu \beta \beta}$. For $f_{2\nu \beta \beta} = 1$, we have a full contribution from the $2\nu \beta \beta$ background, while $f_{2\nu \beta \beta} = 0$ corresponds to a complete reduction of this background. Prospects for depleting $^{136}$Xe is discussed in~\cite{Aalbers:2022dzr}. We do not consider the effect on the Xenon mass number when depleting one of its isotope. Shown are the results using the likelihood in the $S1/S2$ space, and also in the recoil electron energy space, for both Argon and Xenon. The exposure to detect a single component is reduced due to the conversion from ideal E$_r$ space to S1/S2 space. As $\alpha$ increases the test becomes less stringent, consequently the desired exposure $\D\T$ tends to decrease. For the ideal case with a perfect energy resolution and  $f_{2\nu \beta \beta} = 0$, we find that both the pep and CNO signals may be extracted for exposures $\lesssim 100$ ton-years. For a full analysis in the $S1/S2$ space and with $f_{2\nu \beta \beta} = 0$, the exposures for signal identification reach $\gtrsim 10^3$ ton-years. Note that in all cases, relative to other components of the solar flux, the CNO detection significance is more sensitive to the assumed metallicity due to the differences between the high and low metallicity models.

\par
\begin{figure*}[!htbp]
\includegraphics[width = 1.02\textwidth ]{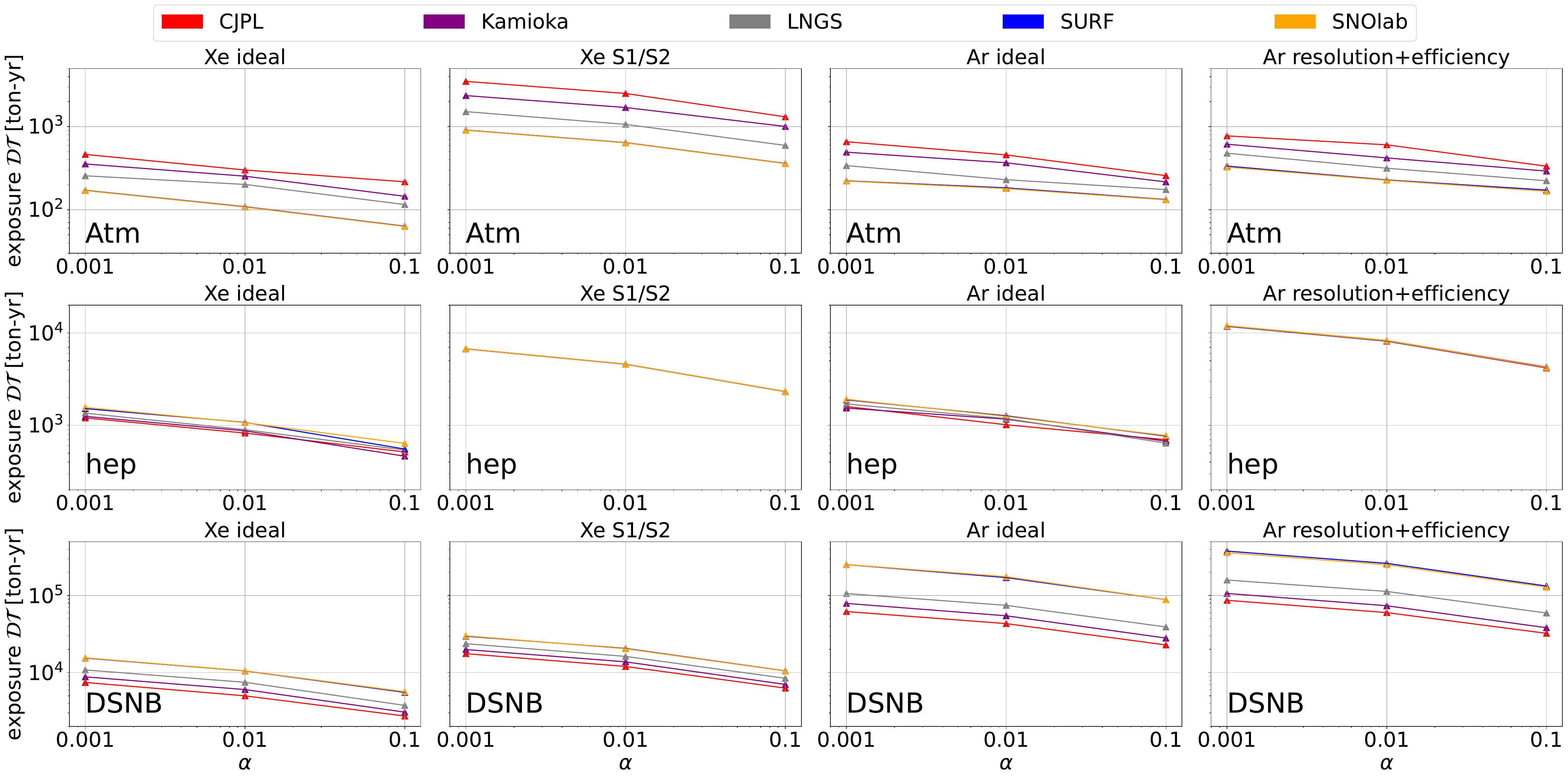}
\caption{Prospects for detection  of atmospheric neutrinos (tow row), hep neutrinos (middle row), and the DSNB (bottom row) in the nuclear recoil channel at several detector locations, for high metallicity, for the given experimental exposures as a function of $\alpha$, as defined in the text. First column shows Xenon ideal, second column shows Xenon $S1/S2$ analysis, third column shows Argon ideal, and fourth column shows Argon resolution plus efficiency model.
}
\label{fig:DTvsALPHA_NR}
\end{figure*}

 We now move onto the nuclear recoil channel. Figure~\ref{fig:DTvsALPHA_NR} shows the detection significances for the hep, DSNB and atmospheric neutrino components for the five detector locations under \mbox{$\alpha$ = 0.001, 0.01, 0.1} and \mbox{$\beta$ = 0.1}, corresponding to 90\% power. For all locations, we compare the ideal cases with no background and perfect efficiency to those in which detector efficiency and backgrounds are modelled. 

\begin{table}[!htbp]

Xenon S1/S2 space $f_{2\nu\beta\beta}$ = 1, high metallicity 
\begin{tabular}{|c|c|c|c|c|c|c|c|c|c|c|}
\hline
$\gamma$ & bg & \multicolumn{3}{c|}{$\alpha$ = 0.001} & \multicolumn{3}{c|}{$\alpha$ = 0.01} & \multicolumn{3}{c|}{$\alpha$ = 0.1}
\\
\hline
& &$\beta$ = 0.2 & $\beta$ = 0.1 & $\beta$ = 0.05 & $\beta$ = 0.2 & $\beta$ = 0.1 & $\beta$ = 0.05 &  $\beta$ = 0.2 & $\beta$ = 0.1 & $\beta$ = 0.05 \\

\hline
\multirow{2}{*}{CNO} & $\kappa_{all, S1/S2}$ - $\gamma$ - $^{85}$Kr - $^{222}$Rn & 8742         & 10813        & 12651         & 5666        & 7349        & 8913         & 2548       & 3717       & 4838        \\

& $\kappa_{all, S1/S2}$ - $\gamma$  & 8903         & 11007        & 12908         & 5760        & 7498        & 9090         & 2598       & 3779       & 4923        \\

\hline
\multirow{2}{*}{pep} & $\kappa_{all, S1/S2}$ - $\gamma$ - $^{85}$Kr - $^{222}$Rn  & 31351        & 38792        & 45493         & 20404       & 26455       & 32054        & 9134       & 13301      & 17423       \\

& $\kappa_{all, S1/S2}$ - $\gamma$ & 31996        & 39552        & 46386         & 20774       & 26944       & 32638        & 9330       & 13597      & 17743       \\

\hline
Atm & \multirow{3}{*}{$\kappa_{all, S1/S2}$ - $\gamma$} & 718          & 910          & 1097          & 514         & 642         & 766          & 249        & 361        & 475         \\

hep & & 5448         & 6741         & 7907          & 3535        & 4589        & 5567         & 1586       & 2321       & 3030        \\

DSNB & & 23960        & 30270        & 35203         & 15629       & 20502       & 24858        & 7072       & 10660      & 13879 \\     

\hline
\end{tabular}
\vskip 5mm 

Argon E$_r$ space resolution+efficiency, high metallicity\\

\begin{tabular}{|c|c|c|c|c|c|c|c|c|c|c|}
\hline
$\gamma$ & bg & \multicolumn{3}{c|}{$\alpha$ = 0.001} & \multicolumn{3}{c|}{$\alpha$ = 0.01} & \multicolumn{3}{c|}{$\alpha$ = 0.1}
\\
\hline
& &$\beta$ = 0.2 & $\beta$ = 0.1 & $\beta$ = 0.05 & $\beta$ = 0.2 & $\beta$ = 0.1 & $\beta$ = 0.05 &  $\beta$ = 0.2 & $\beta$ = 0.1 & $\beta$ = 0.05 \\

\hline
CNO  & \multirow{2}{*}{$\kappa_{all, ER}$ - $\gamma$ + $^{222}$Rn } & 21           & 26           & 31            & 14          & 18          & 22           & 6          & 9          & 12          \\
pep &  & 29           & 36           & 42            & 19          & 25          & 30           & 9          & 12         & 16          \\
\hline
Atm & \multirow{3}{*}{$\kappa_{all, NR}$ - $\gamma$}& 233          & 334          & 428           & 190         & 228         & 316          & 132        & 172        & 213         \\
hep     &  & 33714        & 41976        & 50811         & 21861       & 28986       & 35435        & 9943       & 14649      & 19133       \\
DSNB & & 303435       & 376133       & 443940        & 195696      & 259076      & 316675       & 90155      & 136656     & 173342     \\
\hline
\end{tabular}
\caption{Exposure [ton-yr] for multiple $\beta$ under fixed $\alpha$. Top table shows the Xenon results, and bottom table shows the Argon results. In the second column, the ``-" and ``+" sign in the second column indicates the components that have been removed from and added to $\kappa_{all, S1/S2}$, $\kappa_{all, ER}$ or $\kappa_{all, NR}$. }
\label{tab:fixedalpha}
\end{table}

\begin{figure*}[!htbp]
\includegraphics[width = 0.975\textwidth ]{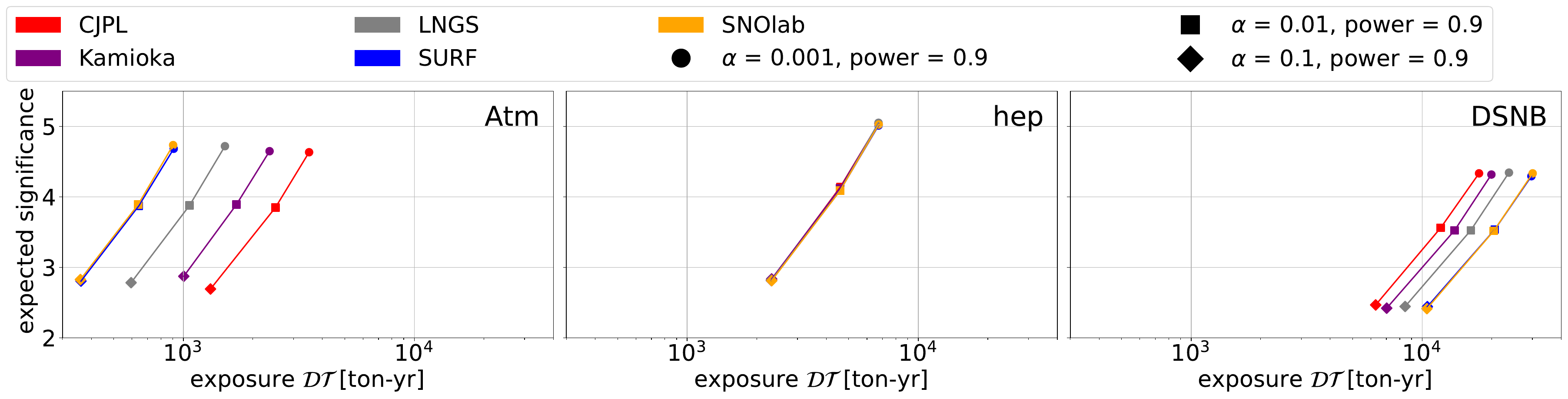}
\caption{Xenon detector in S1/S2 space. Expected significances, as defined in~\cite{2021Jayden}, as a function of exposure for atmospheric, hep, and DSNB neutrinos. The different point types show the optimal exposures as determined in Figure~\ref{fig:DTvsALPHA_NR} for the Type I error rate $\alpha$ = 0.001, 0.01, 0.1 and power = 0.9. 
}
\label{fig:cf_Z_Asimov}
\end{figure*}

\par There are several interesting points to be noted in Figure~\ref{fig:DTvsALPHA_NR}, in particular how the detection prospects change as a function of detector location. For atmospheric neutrinos, the top panel in Figure~\ref{fig:DTvsALPHA_NR} shows how the optimal exposure (the exposure determined by fixing $\alpha$ and $\beta$) changes across the detectors' location; the best detection prospects come from SURF because the flux is the largest at this location, while the most pessimistic detection prospects are from CJPL because of a relatively lower flux. Depending on detector location, background model and the criteria of $\alpha$ and $\beta$, the atmospheric flux will be detectable with anywhere from $\sim 200-2000$ ton-years of exposure. The hep flux is detectable with $\gtrsim 10^3$ ton-yr exposure, while in all cases the DSNB requires larger exposures, $\gtrsim 10^4$ ton-yr. Interestingly, the prospect for detecting the DSNB at CJPL are slightly better than other locations, since the atmospheric flux is the lowest at these locations. Then to detect hep, DARWIN with 40 ton active liquid Xenon would take $\gtrsim 25$ years, and ARGO with 300 ton fiducial mass of Argon would take $\gtrsim 30$ years under current resolution and Argon detector efficiency~\cite{ARGO} For fixed confidence level $\alpha$, different values for the power are shown in Table~\ref{tab:fixedalpha}.

\par  The above results show that the most drastic decrease in  power occurs when including backgrounds and resolution is for the hep and atmospheric components. This can be seen specifically by comparing the Xe ideal to the Xe $S1/S2$ model, and the Ar ideal to the Ar resolution $+$ efficiency model. The hep component is significantly affected by the energy smearing from the ${}^8$B component. On the other hand, the atmospheric significance is strongly affected by leakage from the electron recoil band, due to scattering events from pp neutrinos and from the 2$\nu \beta \beta$ background. Since the DSNB signal in S1/S2 space is concentrated at lower values than atmospheric neutrino signal~\cite{2021Jayden}, the DSNB is less affected by the pp leakage than atmospheric neutrino does. Due to DSNB's low event rate, and its overlapping with atmospheric neutrino, the main background affecting detecting DSNB is the overlapped atmospheric neutrino on it.

\par As discussed above, our statistically methodology introduced in Section~\ref{sec:statistical} is different than previous studies~\cite{2021Jayden,AristizabalSierra:2021kht,Tang:2023xub}, so it is interesting to compare methodologies where possible. These previous studies provided a test statistic and expected significance for atmospheric neutrino signals under different parameters and exposures. For comparison, we also calculated this significance under our optimally determined detector exposures for each of the different solar, atmospheric, and DSNB flux components that we consider. To compute this significance, we first calculated the test statistic defined in~\citet{2021Jayden} as the ratio of the likelihoods under the null and alternative hypothesis. Then we calculated the square root of the test statistic, and the average of this square root over 10,000 simulated data sets is referred to as the expected significance. 
 Before the square root calculation, we replaced the negative value of the test statistic by a zero. The results for each of the fluxes are shown in Figure~\ref{fig:cf_Z_Asimov}. For atmospheric neutrinos, the expected significance generally increases for the high latitude detector locations such as SURF, which implies better prospects for detecting atmospheric neutrinos, and decreases for lower latitude detector locations such as CJPL, which implies worse detection prospects. On the other hand, for the DSNB the expected significance is larger for CJPL than for SURF, implying better detection prospects for the former location. This is the same trend as shown in our primary analysis method.  

\par In addition to the statistical methodology, for comparison to previous studies we show the NEST detector parameters that we use in Table~\ref{tab:detector_file}. 

\begin{table}
\begin{tabular}{c|c|c|c|c|c}
\hline
detector & g1 
 [phd/$\gamma$]  & g2 [phd/e] & Drift Field [V/cm] & Electron Life-time [$\mu$s] & P\_dphe\\
\hline
all enhanced (this work) & 0.3 & 96.922 & 1000 & 5000 & 0.22\\
LZ\footnote{\url{https://github.com/NESTCollaboration/nest/blob/master/include/Detectors/LZ_SR1.hh}} & 0.113569 & 47.067 & 192 & 6500 & 0.214 \\
G3\footnote{\url{https://github.com/NESTCollaboration/nest/blob/master/include/Detectors/Detector_G3.hh}} & 0.118735 & 76.147 & 180 & 850 & 0.2\\
\hline
\end{tabular}
\caption{Detector parameters used for the NEST simulations for this work are shown in the first row. For comparison, we show the detector parameters used in previous LZ experimental analyses and for simulated G3 experiments. }
\label{tab:detector_file}
\end{table}

\section{Conclusions \label{sec:conclusions}}

\par In this paper we have studied the prospects for detecting solar and atmospheric neutrinos in future large-scale Xenon and Argon dark matter detectors. We have in particular focused on how the prospects change as a function of detector location. This is an important consideration which has not been specifically addressed in previous studies, especially because the low-energy atmospheric neutrino flux depends strongly on the detector location. Moreover, we have employed a principled statistical   methodology that allows calculating the detector exposures as a function of the Type-I and Type-II error rates. 

\par Our analysis shows that the best prospects for the detection of the atmospheric neutrino flux are at the SURF location, while  the least pessimistic chances are at CJPL. In addition to the atmospheric neutrino flux, we examine the prospects for detection of the diffuse supernova neutrino background (DSNB) and all components of the solar neutrino flux. For the DSNB, we find that the prospects are best at CJPL, due largely to the reduced atmospheric neutrino background at this location. Since the atmospheric neutrino becomes a major background to detect DSNB in S1/S2 space, the required exposure to detect DSNB is lower at where the atmospheric neutrino flux is lower. The characteristic exposure for significant detection for different combinations of $\alpha$, $\beta$ is around $\sim 10^4$ ton-yr, which is approximately an order of magnitude larger than characteristic exposures discussed in~\cite{Aalbers:2022dzr}. We find that the CNO component of the solar flux is detectable via the electron recoil channel with exposures of $\sim 10^3$ ton-yr for all locations. 

\par For the solar components of our analysis, the uncertainties on the fluxes are well-quantified. However, for the atmospheric and DSNB, the uncertainties are not as well known. For the atmospheric flux, additional systematic uncertainties in addition to the location-dependent uncertainties that we consider arise from the uncertainty on the primary cosmic ray flux, and the structure of the Earth-magnetic field. For the DSNB, uncertainties in the flux prediction arise from the SNII rate, and the SN spectrum for the different neutrino flavors. However, since this flux is not yet detected, it is difficult at this time to gauge systematic uncertainties. 

\par Dark matter detectors provide a unique to measure the neutrino fluxes that we have discussed. The measurements would be complementary to those at future large-scale neutrino detectors. For example, a precise characterization of the solar neutrino flux would allow for novel methods to measure properties of the Sun~\cite{Cerdeno:2017xxl} and neutrino mass matrix parameters via dark matter detectors~\cite{Mishra:2023jlq} and at future detectors such as DUNE, JUNO, and Hyper-Kamiokande~\cite{Brdar:2023ttb}. In addition,  DUNE~\citep{DUNE:2020ypp,Kelly:2023ugn} and JUNO~\citep{JUNO:2021tll,Suliga:2023pve} will be sensitive to primarily charged current channels, and different neutrino energy ranges, for low-energy atmospheric neutrinos. Because of the relatively low event rates for atmospheric neutrinos, even for the large-scale detector that we consider here, a measurement of the event rates at different detectors through different channels would provide an important calibration in order to better understand the systematics in the measurement of the low-energy flux. This is important for the potential extraction of neutrino parameters and new physics from this data~\citep{Kelly:2019itm,Dutta:2020che}. 

\section*{Acknowledgements} 
Y.~Z. and L.~E.~S. are supported by the DOE Grant No. DE-SC0010813.

\bibliography{apssamp}

\end{document}